\documentclass[10pt, conference]{ieeeconf} 

\IEEEoverridecommandlockouts                              

\title{\LARGE \bf
A Self-Negotiation Framework for Ethical Decision-Making during\\Task Interruptions in Service Robots
}

\author{Nele Reichert$^{1}$,  Mashal Afzal Memon$^{2}$,  Marco Autili$^{2}$,  Nico Hochgeschwender$^{1}$
\thanks{$^{1}$Nele Reichert and Nico Hochgeschwender are with the Department of Mathematics and Computer Science at the University of Bremen, Germany
        {reichern@uni-bremen.de}, {nico.hochgeschwender@uni-bremen.de}}%
\thanks{\protect\\ $^{2}$ Mashal Afzal Memon and Marco Autili are with the Department of Information Engineering and Information Sciences and Mathematics, University of L’Aquila, Italy 
        {mashalafzal.memon@univaq.it}, {marco.autili@univaq.it}}%
}

\usepackage{amsmath}
\usepackage{url}
\usepackage{soul}
\usepackage{xcolor}
\usepackage[ruled,linesnumbered]{algorithm2e}
\usepackage{adjustbox}
\usepackage[para]{footmisc}
\usepackage{comment}
\usepackage{algorithmic}
\usepackage{tabularx}
\usepackage{booktabs}
\usepackage{paralist}
\usepackage{longtable}
\usepackage{array}
\usepackage[hidelinks]{hyperref}
\usepackage[capitalise]{cleveref}
\usepackage{booktabs}
\usepackage[table]{xcolor} 
\usepackage{amsfonts}
\usepackage[T1]{fontenc}

\newboolean{showcomments}
\setboolean{showcomments}{true}
\ifthenelse{\boolean{showcomments}}
{\newcommand{\nb}[2]{
  \fcolorbox{black}{yellow}{\bfseries\sffamily\scriptsize#1}
  {\sf\small\textcolor{teal}{\textit{#2}}}
 }
 
}
{\newcommand{\nb}[2]{}
 
}

\usepackage[normalem]{ulem}
\newboolean{showdiffs}
\setboolean{showdiffs}{true}
\ifthenelse{\boolean{showdiffs}}
{
    
    \newcommand\del[1]{\textcolor{red}{\sout{#1}}}

}
{
    
    \newcommand\del[1]{}
}

\begin{document}

\maketitle
\thispagestyle{empty}
\pagestyle{empty}


\begin{abstract}
Service robots operating in public environments frequently encounter interruptions when multiple users request service simultaneously. Resolving such conflicts requires ethical decision-making, as prioritizing one user request can disadvantage another. Current approaches rely on static rules or centralized arbitration and do not support autonomous, ethics-based conflict resolution. This paper addresses the question of how a single robot can arbitrate between multiple users during task interruptions and make ethically aligned decisions without relying on external coordination. We introduce a \emph{self-negotiation framework} that represents each user by an ethical profile that captures their contextual ethical preferences and conditions, and resolves conflicts through an internal negotiation process. The framework is implemented in a modular ROS-based implementation and evaluated in simulation with a realistic interruption scenario. The results show that the system consistently produces user ethical preference-aligned outcomes, supports multilateral negotiation among users, and responds within 1.5 seconds, with near-linear runtime growth under increasing user input.
\end{abstract}


\section{Introduction}
\label{sec:introduction}

\begin{figure}[!htb]
   \centering
    \includegraphics[width=0.8\linewidth]{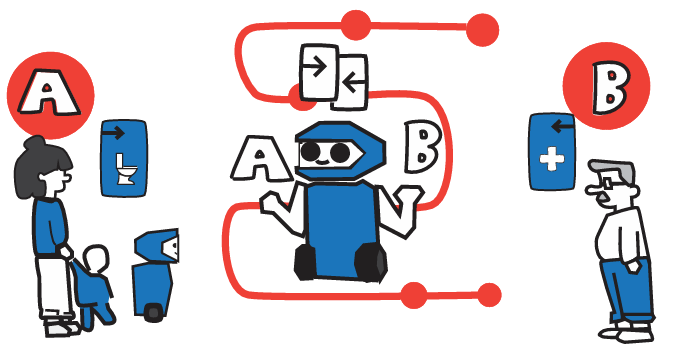}
    \vspace{-8pt} \caption{Interruption scenario: Current user $A$ (Alice, with child) requests guidance to the bathroom, while interrupting user $B$ (Bob, elderly) requests guidance to the pharmacy. The robot must decide whether to continue serving Alice or to interrupt and assist Bob, balancing ethical trade-offs.} 
    \label{fig:scenario}
    \vspace{-8pt}
\end{figure}

\noindent Autonomous robots are increasingly deployed in public spaces such as airports, malls, hospitals, and museums to provide navigation, delivery, or information services. In these dynamic environments, robots already engaged in a task frequently face interruptions by other users. Such situations are common in multi-user human-robot interaction and force the robot to decide whether to continue its current task or respond to the interruption~\cite{Sun2013}. While prior work has examined \emph{when} a robot can appropriately interrupt a human~\cite{Banerjee2018}, our focus is the complementary problem of \emph{ethically handling} human-initiated interruptions. This requires arbitration between competing requests (see~\cref{fig:scenario}), where trade-offs may depend on different contextual needs, e.g., assisting those in greater need, responding to emergencies, or first-come-first-served, though other dimensions may also arise depending on context.
For truly autonomous operation, robots must resolve such conflicts internally rather than relying solely on static rules or external arbitration. Interruption management in service robots has often been approached as a scheduling problem, using fixed priorities or centralized control~\cite{Sun2013}\cite{Dudek2019}\cite{Dudek2020}. These methods are effective for ensuring efficiency and safety, but they do not address the ethical dimension of deciding which user should be prioritized. Moreover, in such interrupting situations, robots designed to support humans should account for the ethical considerations of their users, which can guide decision-making and contextual prioritization. For example, in an airport setting, a user may prefer that an elderly person be prioritized for elevator access when assistive robots are transferring multiple users to different locations.
Research in machine ethics further emphasizes the need to model ethical preferences to manage trade-offs in decision-making~\cite{tolmeijer2020implementations}\cite{alidoosti2022incorporating}\cite{autili2025engineering}, while multi-user HRI highlights privacy concerns since sensitive information should not be disclosed across users~\cite{reig2020}.

\vspace{-10pt}

These challenges show that interruption management cannot be reduced to efficiency alone. Our work addresses this gap by introducing a \emph{self-negotiation framework} that enables a robot to arbitrate ethically between simultaneous user requests. 
While prior work has explored negotiation mechanisms for coordinating multiple autonomous agents, such approaches typically do not consider ethical trade-offs~\cite{kiruthika2020lifecycle}. In the limited cases where ethical considerations are incorporated~\cite{memon2025robethichor}, they are addressed through two context-dependent robots acting independently on behalf of individual users, often without accounting for scenarios involving more than two users. Moreover, these approaches generally assume that robots pursue their own tasks and do not consider situations in which a robot must interrupt an ongoing task. In contrast, we address a fundamentally different and unexplored challenge: how a single robot can negotiate internally among the ethical perspectives of multiple users. 
This intra-robot setting raises unique requirements, including modeling several users within one agent, integrating goal-dependent interruptibility, preserving privacy without external coordination, and supporting multilateral conflicts beyond bilateral cases. We therefore ask: \emph{How can a single robot autonomously and ethically arbitrate between conflicting user requests during task interruptions without external coordination or data disclosure?} To answer this, we propose a \emph{self-negotiation framework} realized in a modular ROS-based architecture and evaluated in simulation. 
Our work follows the concept of Floridi~\cite{floridi2018soft}, which defines hard ethics as laws and regulations and soft ethics as user ethical preferences. Hence, the robots in our work comply with laws and make decisions based on user ethical preferences unless they violate the law.
%

The main contributions of our work are: (i) introducing the concept of \emph{self-negotiation}, enabling a single robot to arbitrate internally between multiple users’ ethical profiles; (ii) formalizing task interruptions in multi-user service robots as an ethical decision-making problem, capturing user dispositions, statuses, and context-dependent priorities; (iii) realizing the framework in a modular ROS-based architecture that models multiple users, integrates goal-dependent interruptibility, and supports multilateral negotiation; and (iv) evaluating the framework in simulation, showing that it produces consistent, preference-aligned outcomes and maintains real-time performance as user complexity increases.

\vspace{4pt}
The structure of the paper is as follows. Section \ref{sec:related} discusses the related work. A motivating scenario is presented in Section \ref{sec:motivation}. The self-negotiation framework is detailed in Section \ref{sec:self_negotiation}. An architecture for implementing the framework is presented in Section \ref{sec:architecture} and evaluated in Section \ref{sec:experiments}. Finally, limitations and future work are discussed in Section \ref{sec:discussion} and a conclusion is drawn in Section \ref{sec:conclusion}.

\section{Background and Related Work}
\label{sec:related}

\noindent We review related work on ethical decision-making, task interruptions, and negotiation.

\subsection{Ethical Decision-Making in Autonomous Systems} 
\noindent Integrating ethics into autonomous decision-making remains challenging. Abstract ethical theories are difficult to operationalize in robotic control, while user values are diverse and context-dependent~\cite{moor2006nature}\cite{allen2006machine}. One approach encodes ethical theories into control, operationalizing Asimov's laws in planners that simulate outcomes to select ethical options~\cite{Winfield2019}. Although these approaches show robots can follow principles, they remain detached from human values. Another approach embeds user-specific ethical preferences to align with individual priorities~\cite{alidoosti2022incorporating}\cite{autili2025engineering}\cite{autili2025reference}. 
However, these values vary by context~\cite{tolmeijer2020implementations}. Recent works distinguish between hard ethics and soft ethics~\cite{floridi2018soft}. Prior research shows ethical reasoning's importance but does not address arbitrating between competing user ethical preferences.

\subsection{Task Interruptions in Service Robots}
\noindent Service robots in public spaces face task interruptions when approached by new users while assisting others~\cite{Sun2013}\cite{eklundh2003}. Much literature treats this as an optimization problem in task allocation, focusing on efficiency~\cite{Sun2013}\cite{Dudek2019}\cite{Dudek2020}\cite{Haigh1998}. Studies have addressed technical issues like safe interruption handling~\cite{Dudek2020} and perception mechanisms for detecting interruptions~\cite{Kabir2018}. User studies show that acceptance of interruptions improves when framed as benefiting another user~\cite{Carter2022}\cite{Carter2024}. Practical aspects, such as detour length, affect user perception~\cite{Carter2022}. However, these works emphasize efficiency and safety while neglecting ethical trade-offs. 

\subsection{Negotiation-Based Approaches}
\noindent Automated negotiation has been widely studied for autonomous agents to resolve conflicts and reach agreement 
~\cite{kiruthika2020lifecycle}\cite{baarslag2017automated}\cite{memon2025systematic}. Negotiation protocols define interaction rules, from simple one-shot offers, where one party proposes and the other accepts or rejects~\cite{ji2014one}, to alternating-offer strategies where agents iteratively refine proposals~\cite{aydougan2017alternating}. Complex settings extend negotiations to multilateral scenarios involving multiple agents~\cite{caillere2016multiagent}. While negotiation is common in economics~\cite{li2003review}\cite{baarslag2017value}, where preferences are monetary, it has been applied in privacy-sensitive domains, such as bargaining over user data~\cite{baarslag2017automated}\cite{filipczuk2022automated}. The study in~\cite{memon2025robethichor}
introduced negotiation over users' ethical preferences, enabling robots to decide which one should be prioritized in shared-resource conflicts (e.g., elevator usage). However, the approach is limited to pairwise interactions and does not account for scenarios involving more than two users, which motivates the need for negotiation approaches that can handle multiple users’ ethical preferences simultaneously. 

\section{Motivation: Interruption Scenario}
\label{sec:motivation}
\noindent We use the scenario in~\cref{fig:scenario} to illustrate 
the need of ethical decision-making in service robots based on users ethical preferences. Users provide ethical profiles and their contextual status\footnote{Users voluntarily provide contextual information and may choose to withhold/opt-out certain details. In such cases, the robot does not incorporate the omitted information into the negotiation process, which may affect the resulting service outcome. Moreover, the robot operates with predefined emergency categories: some require external verification (e.g., boarding pass scanning to confirm a flight-related emergency), while others are inferred through sensing. This design limits the possibility of strategic misrepresentation and deception, beyond the assumption of a cooperative environment in which users trust the robot.} to robots through an app developed by the service robot provider, to customize its decision-making. Although users can customize the robot’s decisions based on soft ethics (i..e, their contextual ethical preferences), the robot will always comply with hard ethics (i.e., laws and regulations). Once the interaction concludes, the robot refreshes its state and does not retain any personal information. Similar to real-world settings, a single robot may be interrupted by multiple users and therefore has the capabilities of arbitrating among multiple user profiles. Our case study targets one specific context; however, such robots can be deployed across a wide range of real-world environments, and hence our approach can be generalized to diverse scenarios and contexts.


\subsection{Scenario}
\noindent Our scenario shows a dilemma in public spaces: a robot must arbitrate between competing requests with ethical implications. Here, urgency (Alice with a child needing a bathroom) conflicts with accessibility (Bob as elderly), as shown in~\cref{fig:scenario}. Service robots serve registered customers who provide ethical profiles via an application. Requests include situational conditions, e.g., being in an emergency, distress, fatigue, or accessibility needs. Some conditions are explicitly provided by users, while others can be inferred from data or sensors. In the case of an interruption, the robot must decide which request to prioritize while ensuring users cannot learn about others' conditions. This setting provides the basis for our self-negotiation framework, where ethical profiles capture user dispositions.

\subsection{Ethical Profiles}
\noindent Building on the intuition in~\cite{dennis2016formal}\cite{Nallur:2020}\cite{bremner2019proactive}, where (i) users' ethical preferences are modeled as soft constraints rather than hard vetoes on tasks, and (ii) proposes the explicit operationalization of ethical concepts through the encoding of values for utility-based reasoning, we propose the generation of ethical profiles in our approach. An ethical profile is defined as a structured representation of a user’s contextual moral preferences. It contains a set of ethical dispositions associated with contextual ranks that indicate the relative priority a user assigns to specific ethical considerations within a given context, as summarized in Table~\ref{tab:dispositions_ranks}.

 Profiles can be provided: (i) explicitly during registration, (ii) initialized from normative defaults and adapted over time, or (iii) learned from user feedback. Some situational conditions can be inferred by the robot using sensors or background information.
Formally, let $\mathcal{D} = \{ d_1, \ldots, d_m \}$ be a set of dispositions with $m > 0$, and $rank() : \mathcal{D} \times \mathcal{C} \rightarrow \mathbb{N}$ a function that assigns ranks to dispositions in a context $\mathcal{C}$. 

An ethical profile is a pair $E_\pi(\mathcal{D},\mathcal{R})$, where $\mathcal{R} = \{\leq_{C_1}, \ldots, \leq_{C_l}\}$ with $l > 0$ being a set of context-dependent non-strict partial order relations on $\mathcal{D}$. The ethical profile for a specific context $\mathcal{C}_i$ with $\leq_{\mathcal{C}_i}\in \mathcal{R}$ is a partially ordered pair $E_\pi(\mathcal{C}_i) = (\mathcal{D},\leq_{\mathcal{C}_i})$. The relations are constructed such that for a pair of dispositions $(d_{j},d_{k}) \in\leq_{\mathcal{C}_i}$ with $1 \leq j \leq m$ and $1 \leq k \leq m$, it holds that $rank(d_{j},\mathcal{C}_{i}) \leq rank(d_{k},\mathcal{C}_i)$. Table~\ref{tab:dispositions_ranks} shows the profiles of two users in our scenario. A range of $[1,5]$ is fixed for the ranks. Alice, in their contextual ethical profile, prioritizes giving high precedence to people with health conditions with rank $5$, to anyone accompanying a small child with rank $4$, and so on. Bob, in their profile, emphasizes giving high precedence to anyone accompanying a small child with rank $5$, followed by precedence to an elderly person with rank $4$, and so on. Following these values, the ethical profiles for Alice in this context would be $E_\pi^A(C^A) = (\mathcal{D}, \leq_{C^A})$ with $d_2 \leq_{C^A} d_3 \leq_{C^A} d_4 \leq_{C^A} d_6 \leq_{C^A} d_1 \leq_{C^A} d_5$ and the ethical profile for Bob would be $E_\pi^B(C^B) = (\mathcal{D}, \leq_{C^B})$ with $d_6 \leq_{C^B} d_3 \leq_{C^B} d_4 \leq_{C^B} d_5 \leq_{C^B} d_2 \leq_{C^B} d_1$.
The framework generalizes to arbitrary sets of dispositions in different domains and supports multilateral negotiations beyond bilateral cases.

\begin{table}[ht]
    \centering
    \rowcolors{2}{white}{gray!10} 
    \begin{tabular}{>{\columncolor{gray!20}}c l c c }
        \toprule
        Id & Disposition: Give precedence to \dots & Alice & Bob  \\
        \midrule
        $d_1$ & people with a small child        & 4 & 5  \\
        $d_2$ & elderly people                   & 2 & 4  \\
        $d_3$ & disabled people                  & 2 & 3  \\
        $d_4$ & people in urgency                & 3 & 3  \\
        $d_5$ & people with health conditions    & 5 & 3  \\
        $d_6$ & people with crowd anxiety        & 3 & 1  \\
        \bottomrule
    \end{tabular}
    \caption{Exemplary ranks of two users over ethical dispositions.}
    \label{tab:dispositions_ranks}
    \vspace{-3em}
\end{table}

\subsection{Context and User Status}
\noindent When making a request, users are associated with two kinds of information: their situational \emph{status} and the surrounding \emph{context}. The \emph{user status} captures conditions specific to an individual, such as urgency or disability. It is modeled as a set of pairs $\mathcal{S} = \{\langle c_1, v_1 \rangle, \ldots, \langle c_p, v_p \rangle\}, v_i \in \{\text{true}, \text{false}\},$
where each $c_i$ denotes a possible condition and $v_i$ indicates whether it applies.  The \emph{context} describes external factors such as location or environment. It is modeled as a set of attribute–value pairs
$\mathcal{C} = \{\langle a_1, v_1 \rangle, \ldots, \langle a_n, v_n \rangle\}, \quad n > 0,$
with $a_i$ as an attribute name and $v_i$ its value. Context can influence user status, since some conditions are not explicitly declared but derived from environmental attributes. 
In our scenario, Alice’s status activates disposition for \emph{with child} ($d_1$) and \emph{urgency} ($d_4$), while Bob activates \emph{elderly} ($d_2$). 
The context (e.g., ``location = mall'') provides background information but remains distinct from user-specific statuses.


\subsection{Decision-Making}
\noindent The robot arbitrates between conflicting requests based on users' ethical profiles and statuses by assessing the ethical impact of serving one user over another and determining request precedence. In our example, both profiles agree that Alice's conditions (urgency, small child) outweigh Bob's needing the service due to being an elderly person, so the robot continues assisting Alice while deferring Bob's request. More generally, the problem is deriving a precedence relation among users in a privacy-preserving way. The next section develops this into a self-negotiation framework that formalizes the decision process.

\section{Self-Negotiation Framework}
\label{sec:self_negotiation}

\noindent Ethical decision-making in multi-user service robots requires arbitration between conflicting user requests. We conceptualize this as \emph{self-negotiation}. In classical negotiation, agents exchange and evaluate offers from their perspectives. In self-negotiation, a robot generates and evaluates offers for its users, alternating internally between their perspectives. This preserves the negotiation structure, offers, counter-offers, acceptance or rejection, while adapting it to an intra-robot setting without external communication. 
When evaluating offers, the negotiation process in our framework avoids reducing the problem to a one-shot global optimization over complete ranked lists. Instead, it evaluates context-activated conditions within a structured negotiation procedure, supporting incremental reasoning and explainable arbitration.

\subsection{Goal-Task-Action Model}
\noindent We adopt a goal--task--action decomposition, where goals are refined into tasks and executable actions. A navigation goal $\langle g_1 = \text{navigate to bathroom\_1} \rangle$ may decompose into tasks like $\langle t_3 = \text{call elevator} \rangle$, which require actions like $\langle a_1 = \text{press elevator button} \rangle$. This hierarchy aligns with established behavior specifications in robotics, such as behavior trees~\cite{btree} or task planners, making it compatible with common control architectures. In RobEthiChor, negotiation occurred at the task level, with each robot's goal fixed. However, interruptions in a single service robot require arbitration at the \emph{goal level}: deciding which user's goal to pursue~\cite{memon2025robethichor}. Switching goals leads to new tasks and actions, but the key choice is which user's goal takes precedence. This shift is essential for handling interruptibility, motivating the notational changes in the following formalization. We consider $\mathcal{G}$ as the set of all goals specified in the system. 

\subsection{Ethical Evaluation}

\noindent To arbitrate between users, the robot must quantify how ethical decisions appear from each user's perspective. Building on ethical profiles introduced earlier, we define two concepts: \emph{ethical implication} and \emph{ethical impact}. 
The \emph{ethical implication} $E_\pi(g,\mathcal{C}) \subseteq \mathcal{D}$ of a goal $g \in \mathcal{G}$ specifies dispositions affected by pursuing $g$ in context $\mathcal{C}$. These sets are defined at design stage by analyzing how goals and tasks influence users. At runtime, the implication is restricted by active user status $E_\pi(g,\mathcal{C},\mathcal{S}) \subseteq E_\pi(g,\mathcal{C})$,
including only dispositions triggered by the current user conditions.
The \emph{ethical impact} assigns a numerical score to a goal from the user's perspective $
\mathcal{G_{EI}}(g,\mathcal{C},\mathcal{S}) = \sum_{d \in E_\pi(g,\mathcal{C},\mathcal{S})} rank(d,\mathcal{C})$.
This aggregates how strongly relevant dispositions are prioritized in a user's ethical profile. In the mall scenario, Alice's profile prioritizes giving precedence to people
with health conditions, while Bob emphasizes giving precedence to anyone accompanied by a child. If Alice requests bathroom guidance and Bob requests pharmacy assistance, ethical impacts reflect these contrasting priorities.
In self-negotiation, the robot alternates perspectives: generating offers based on one user's status and evaluating them from another's profile by computing ethical impact. 

\subsection{Negotiation Protocol}
\noindent In our framework, we adopt the alternating-offer protocol 
In its classical form, negotiation proceeds in rounds: one agent (sender) proposes an offer, the other (receiver) evaluates and either accepts it or responds with a counteroffer, after which the roles are switched. This simple back-and-forth structure provides a foundation for our self-negotiation framework.

In our framework, an offer represents a user’s goal together with the subset of 
conditions they are willing to reveal. A \emph{complete offer} is defined as
$\bar{o} = \langle g, \mathcal{S}_{\bowtie} \rangle$,
where $g \in \mathcal{G}$ is the user’s goal and 
$\mathcal{S}_{\bowtie} \subseteq \mathcal{S}$ contains only the conditions that are 
true for the user. Negotiation does not reveal all of this information at once. Instead, the robot constructs 
\emph{partial offers} incrementally, following a minimal strategy in which exactly one additional condition is disclosed per round:
$o_i = \langle g, \mathcal{S}_i \rangle , \quad \mathcal{S}_i \subseteq \mathcal{S}_{\bowtie}, \ |\mathcal{S}_i| = i$.
This stepwise revelation mirrors classical negotiation dynamics, while preserving privacy 
by limiting information exposure.

An offer should be accepted by the receiver if it can be considered to be more ethical to the receiver than their own complete offer. This is expressed through a utility function $\mathcal{U}(o_i^S,\bar{o}^R) = \mathcal{O_{EI}}(o_i^S,\mathcal{C}^R) - \mathcal{O_{EI}}(\bar{o}^R,\mathcal{C}^R)$ calculating the ethical impact of the sender's partial offer $o_i^S$ and the receiver's complete offer $\bar{o}^R$, both from the perspective of the receiver. 
The acceptance criteria is formally defined as:
$$\mathcal{N}(o^S_i,\bar{o}^R)=
\begin{cases}
accept, & \text{if }\mathcal{U}(o^S_i,\bar{o}^R)>0\\
reject, & \text{if }\mathcal{U}(o^S_i,\bar{o}^R)\leq0 \text{ and }i <  |O^R| \\
quit & \text{otherwise }
\end{cases}$$

If an offer is accepted, the negotiation terminates with an agreement that both users would consider the most ethical outcome in the given context. Otherwise, the offer is rejected, the receiver issues a counter-offer revealing one additional condition, and the roles are switched again. A user who has no further conditions to reveal \emph{quits} 
the negotiation. Upon quitting (i.e., when negotiation ends without agreement), the robot's mission does not fail. In such cases, the robot applies default built-in rules informed by hard ethics, e.g., a first-come-first-served policy. Moreover, at any time, the negotiation can be overruled by emergency events, if the request interruption request falls under hard ethics, leading the system to follow its built-in behavior. 


In self-negotiation, the alternating offer protocol’s structure remains the same, but offers are not exchanged between separate agents. Instead, the robot alternates perspectives by generating offers based on one user’s status and evaluating them using the other user’s ethical profile. This requires access to both users’ data and careful switching of roles in each round. In the interruption scenario, the \emph{interrupting user} always starts as the sender, whereas the \emph{currently active user} begins as the receiver. Roles are then alternated in subsequent rounds, as in the classical alternating offer protocol.

\section{Architecture Overview}
\label{sec:architecture}
\begin{figure}[!htb]
   \centering
    \includegraphics[width=1\linewidth]{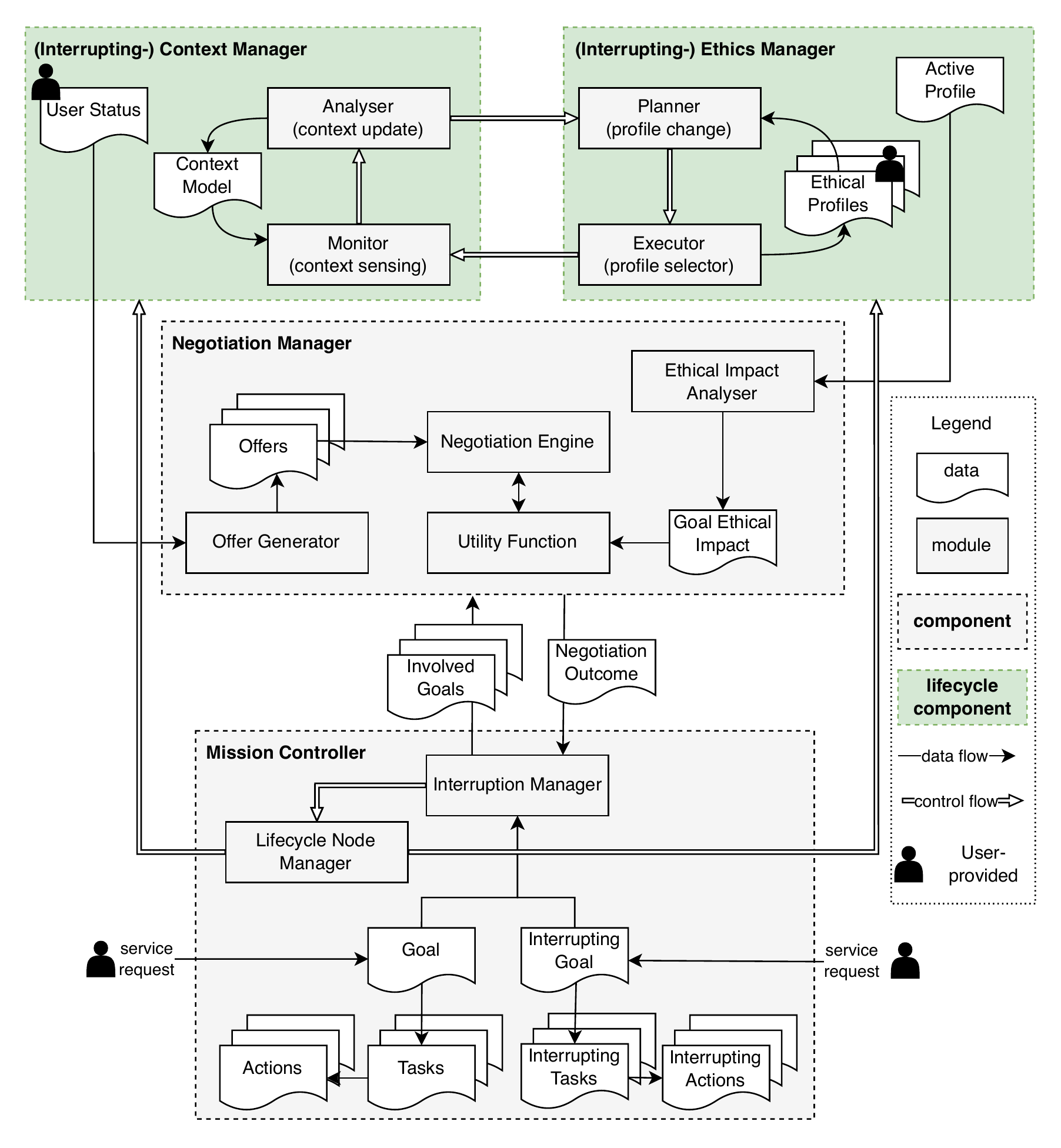}
      \caption{Architecture of the system displaying the four main components, their submodules, and their control and data flow.}
    \label{fig:arch}
    \vspace{-2em}
\end{figure}

\noindent While the previous section introduced self-negotiation theory, this section describes how these concepts are embedded in a robot control architecture. 
Embedding self-negotiation into a service robot requires more than an abstract decision model; it needs an architecture integrating ethical reasoning with the robot's control system while meeting real-time and privacy requirements. We designed a modular ROS~2 architecture that operationalizes the concepts introduced in~\ref{sec:self_negotiation}. \cref{fig:arch} provides an overview of the proposed architecture. 

The architecture links self-negotiation framework elements to system components: the \emph{Context \& Ethics Manager} maintains ethical profiles and user statuses; the \emph{Negotiation Manager} implements the alternating-offer protocol for intra-robot arbitration; and the \emph{Mission Controller} manages goals and interruptibility within the goal–task–action hierarchy. A key novelty lies in using lifecycle nodes to preserve privacy: user instances exist only during negotiation and terminate afterwards, ensuring sensitive data are not permanently stored. 

\subsection{Context and Ethics Manager}
\noindent The \emph{context manager} and \emph{ethics manager} manage the ethical profile, user status, and context of specific users.
During the interruption, the data of both users must be maintained within the same robot. To ensure separation, the context and ethics manager tracks their data in separate instances, identical in function but working on different datasets. This ensures clear data ownership and prevents accidental leaks. This doubling of instances is not visible in the architectural design, as both instances interact with the system identically. 
The interruption scenario requires dynamic data management. The interrupting user's data enter the system during an active mission and should leave once negotiation ends to keep private data in the system only when necessary. 
To achieve this, the context and ethics manager is adapted into \emph{lifecycle components}, which can be activated and deactivated. When active, they receive, hold, and update user data. When deactivated, all data are deleted and configurations reset.
This allows components to react to user encounters while minimizing private data storage duration. 
The lifecycle components concept extends to current user data. 
While idle, both sets of context and ethics managers remain deactivated. When starting a service, the corresponding managers activate to track user data.
During interruption, the second set activates for the interrupting user. After negotiation, the managers for users without precedence deactivate, removing their data. 
Data for users with precedence remain until mission completion, after which their manager deactivates and the robot returns idle.
The context and ethics manager also controls the cycle between the \emph{planner}, \emph{executor}, \emph{analyzer}, and \emph{monitor} modules, updating the ethical profile based on context and selecting active profiles for negotiation.

\vspace{-0.8em}
\subsection{Mission Controller}
\noindent The \emph{mission controller} receives user goals and processes them into tasks and actions executable by the robot. 
All functionalities needed to handle interruption are managed by the \emph{interruption manager} and \emph{lifecycle node manager} modules, which are new system additions. 
The interruption manager handles pre- and post-processing of interruption and negotiation requests. Upon receiving an interruption, it signals the lifecycle node manager to activate the interrupting user's context and ethics manager. The interruption manager tracks their start-up process, as they must be ready to provide data before negotiation begins. 
Once ready, the interruption manager sends a request to the negotiation manager. After the negotiation ends, it processes the outcome for needed changes, such as changing the robot's active goal if the interrupting user got precedence. It signals the lifecycle node manager to deactivate the second context and ethics manager.


\subsection{Negotiation Manager}
\noindent The \emph{negotiation manager} executes the negotiation protocol when prompted by a negotiation request from the mission controller, which specifies the goals to be negotiated.
It represents two perspectives between which negotiation occurs and switches between them as user roles alternate. 
Two instances of the \emph{offer generator} and \emph{ethical impact analyzer} exist in the negotiation manager, one per user. 
The offer generator queries the context manager of its appointed user regarding their status and creates offers based on specified goals. The ethical impact analyzer calculates the ethical impact of offers based on the active profile of its assigned user from the respective ethics manager. 

The \emph{negotiation engine} and \emph{utility function} implement the main algorithm logic. The negotiation engine coordinates the alternating protocol by generating and evaluating offers, switching roles, and terminating negotiation when an agreement is reached or no offers remain. The utility function evaluates offers from a specific user's perspective, as directed by the negotiation engine. Both must switch between correct instances of the offer generator and ethical impact analyzer based on role assignment. 
The negotiation outcome is returned to the mission controller.

\section{Experimental Evaluation}
\label{sec:experiments}
\noindent To evaluate the proposed self-negotiation framework, we focus on two research questions:
\begin{itemize}
     \item \textbf{RQ1 (Correctness):} Does the framework consistently find ethical agreements between conflicting user requests that align with the expected outcomes? 
    \item \textbf{RQ2 (Scalability):} How does the runtime of the negotiation process scale with increasing numbers of user dispositions and active conditions? 
    
\end{itemize}

\begin{figure*}[!htb]
   \centering
    \includegraphics[width=1\linewidth]{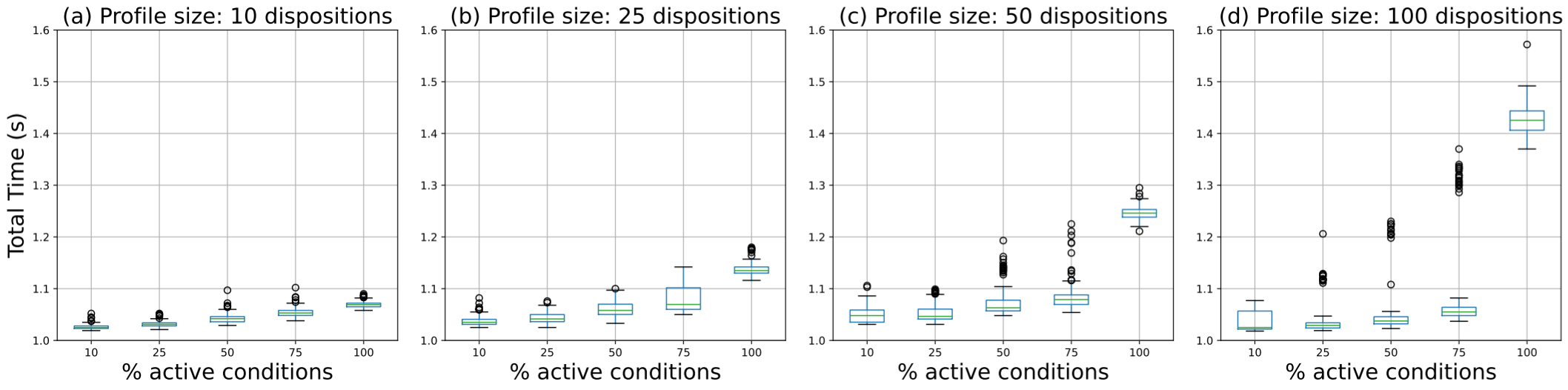}
    \caption{Comparison of the runtime with different numbers of conditions and dispositions. Each plot has a fixed number of dispositions, while the x-axis specifies the percentage of activated user conditions.}
    \label{fig:exp}
    \vspace{-0.45cm}
\end{figure*}


\subsection{Experimental Setup} 
\noindent Our framework was implemented in ROS~2 Humble and deployed on a laptop with AMD Ryzen~7 CPU, Nvidia RTX~4060 GPU, and 32~GB RAM running Ubuntu~22.04. The system was integrated with the TIAGo robot in a Gazebo simulator. Gazebo visualized decisions: the robot was placed in front of two rooms, each assigned to a user request, so that navigation revealed the arbitration outcome. The experimentation is available in the replication package.\footnote{\url{https://bitly.cx/84ll}} 

\subsection{RQ1: Correctness}
\noindent To validate correctness, we first manually constructed a ground truth by determining the expected negotiation outcome for each possible pair of users under both role assignments, based on their profiles and statuses. We then executed the negotiations automatically within the framework.

We generated $10$ fictional users with diverse conditions and ethical preferences. Each user profile included up to $10$ possible conditions and $6$ dispositions reflecting their contextual ethical priorities. Table~\ref{tab:dispositions_ranks} presents exemplary disposition rankings for a pair of users. For each of the $45$ unique user pairs, we computed the expected outcome under both role configurations (i.e., either user being currently served while the other interrupts), resulting in $90$ ground-truth cases.

The logs obtained after running the simulations were analyzed and compared with the manually established ground truth to assess whether the system’s behavior corresponded to the expected outcomes and to measure the time required to complete each negotiation. In all $90$ cases, the outcomes produced by the framework fully matched the ground-truth results and were consistent with the ethical preferences defined in the users' profiles. Moreover, repeated runs involving the same pair of users always produced identical outcomes, regardless of the order, i.e., whether a user started as the interrupted(receiver agent) or the interrupting(sender agent).
Thus, for each pair, the self-negotiation mechanism reliably identified the ethically preferred decision.

\vspace{-0.5em}
\subsection{RQ2: Scalability}
\noindent To assess scalability, we automatically generated test cases with controlled numbers of dispositions and conditions and randomly generated ranks. Profiles were created with 10, 25, 50, and 100 dispositions for each profile. The upper bound of 100 dispositions represents a conservative stress test, as in practice, each disposition must be explicitly ranked by a user, and such a high number is unlikely. For each disposition, one user condition was created, and proportions of 10\%, 25\%, 50\%, 75\%, or 100\% were marked as active. As only active conditions contribute to the offers, this factor directly influences the size of the negotiation space. A 100\% activation rate corresponds to the worst-case runtime, but also an unrealistic case where agreements are unlikely because all offers carry the same weight. 

The two factors were scaled separately by fixing one at its maximum value and varying the other. For each parameter combination, ten ethical profiles and ten user statuses were generated, yielding 90 test cases per combination. All experiments were performed without Gazebo integration to isolate the runtime performance. 
The results are shown in~\cref{fig:exp}. The effect of the number of active conditions can be seen within each plot, whereas the effect of the number of dispositions appears across plots. The runtime increases with both factors, from $\sim$ 1.0s for the smallest cases (10 dispositions, 10\% active conditions) to 1.4--1.6s for the largest cases (100 dispositions, 100\% active conditions), with a few outliers. These outliers correspond to instances in which a larger number of active dispositions produce competing constraints during the internal negotiation process, requiring additional negotiation iterations before a consistent decision is reached. Although the runtime increases with the input size, the overall decision time remains within approximately 1–2 seconds, which is acceptable for real-time decision-making in public human–robot interaction scenarios. 

It is important to note that these values represent the complete process, from receiving the interruption to finishing the negotiation. This includes the preparation overhead for initializing the lifecycle nodes of the interrupting user. 
This initialization overhead accounts for approximately 1s of the total runtime. The results are shown in Fig.~\ref{fig:exp}.

\section{Discussion}
\label{sec:discussion}
\noindent We discuss the limitations of the presented approach as well as some areas for future work. 

\subsection{Ethics Statement}



\noindent Machine ethics is broad and encompasses many different challenges.
As far as our approach is concerned, we do not claim to offer a universal solution guaranteeing that a robot will always behave ethically in accordance with universally accepted moral values. We can assume the technical feasibility of encoding ethical principles to embed context-dependent values into machines through dispositions (over which users retain control). However, we cannot assume that our approach will be regarded as ``acceptable'' across diverse fields such as psychology, philosophy, and social science.  
Our system considers user ethical preferences without enforcing commitment to a single ethical theory. Our negotiation strategy's ethical utility function is inspired by utilitarian ethics~\cite{mill2016utilitarianism}. 
Utilitarianism evaluates actions in terms of outcomes, and an action is morally right if it promotes the greatest good for the greatest number~\cite{moor2006nature}. In our system, user-specified ethical profiles with ranked dispositions implement this principle by letting an individual decide how the robot ought to act according to their values and preferences in variable contexts. This enables the system to dynamically adjust and negotiate actions in a manner compatible with the utilitarian goal of optimizing outcomes with respect to user-supplied utility metrics. Moreover, it is important to note that the ethical preferences and ranks provided in a user’s profile are applied universally, i.e., if a user prioritizes a particular ethical consideration (e.g., emergency precedence), the system applies that ranking to all users equally when calculating the precedence of emergencies from that user’s perspective, rather than privileging the individual alone. Thus, the evaluative criteria are not self-privileging but operate under a unified standard across stakeholders. Moreover, this ensures that the robot’s decision-making process does not amount to ethical egoism or altruist behavior, but instead remains aligned with a consequence-based utilitarian framework.

\emph{Implicit Ethics.} A main goal of the system is to provide users with explicit agency regarding robot ethics. However, designers may unconsciously embed their own ethical beliefs into the system, for example, when choosing ethical dispositions or defining ethical implications. This could prevent users from specifying ethical profiles that meaningfully represent their preferences, threatening the system's internal validity. To minimize this effect, defining these ethical dimensions should not be left solely to robot software engineers but should instead be an interdisciplinary effort involving ethicists or user groups to determine how these factors should be designed~\cite{araiza2025roadmap}.

\emph{Privacy Considerations.}
As the framework uses sensitive data, concerning personal values of users, it is designed to ensure no personal information in the negotiation is revealed to other users. The data are only held in the robot's storage as necessary and destroyed once the negotiation or mission ends. For adaptation to self-negotiation, robots must hold complete data of two users simultaneously, raising privacy concerns. Throughout the process, care was taken to separate users' data to prevent data leaks or mix-ups.

\vspace{-0.5em}
\subsection{Simulation}
\noindent The framework has been evaluated in simulation using ROS2-Gazebo setup and manually crafted user profiles. While this may limit external validity, simulation-based testing is standard in robotics, enabling controlled experiments before physical deployment. The architecture was designed for direct transfer to real robots through ROS integration.
Using fictional user data provides a testbed for validating correctness and scalability. Although real user data would capture richer ethical preferences, our controlled setup was necessary to establish ground truth and test negotiation outcomes. This evaluation confirmed that the framework reached agreements aligned with specified profiles and scaled efficiently.
A natural continuation is to complement simulation with user studies. Collecting real ethical profiles would test how people express priorities and perceive robots that arbitrate according to these values. Such studies could reveal mismatches between stated preferences and experiences, offering insights into human–robot value alignment. Deploying the system on physical robots in public settings will be essential for testing robustness under real-world uncertainties. To the best of our knowledge, there are currently no existing approaches that address intra-robot, ethics-based self-negotiation under context-activated user profiles; therefore, a direct comparison with alternative methods was not performed. However, we compare our negotiation mechanism with prior work that considers negotiation among two robots with two different goal-task-action models, unlike ours. The evaluation indicates that our proposed negotiation mechanism exhibits lower runtime overhead and achieves faster resolution.
The proposed self-negotiation framework and interruption scenario open up many implications for expanding the decision-making process.

\emph{Services, Tasks, Robot Capabilities.}
The offered service was considered to serve a single user that the robot could fulfill independently. In a real setting, the robot would offer multiple services, introducing challenges such as serving multiple users or requiring assistance from another robot or human worker. 
Accounting for different service types and modeling them in detail would reveal additional nuances in the decision process. 
Future work could focus on modeling service requirements and robot capabilities to verify whether a robot can take on new requests before negotiation.
Additionally, tasks were assumed to be interruptible. In practice, it would be necessary to model and monitor task interruptibility to account for safety-critical tasks or rollback behaviors.


\emph{Robot Fleet.} In our work, the robot is considered a single entity. However, it should be integrated into a fleet of service robots in a shared environment that can communicate with each other. A robot can then call other robots to serve users who did not receive precedence. This can improve the user experience, as their service request is not rejected but transferred to a different robot.
However, this would pose additional challenges, requiring a policy on which robot should pick up the second goal, based on the robot's distance to the user, whether it is idle, or a combination of factors.

\emph{Multi-lateral Negotiation.} Although a bilateral self-negotiation framework was presented, the framework enables multilateral negotiations by executing multiple bilateral negotiations and designing a policy to combine results. Such a policy requires careful consideration - for example, determining the ethical course if several interrupting users agree one should get precedence, but the active user disagrees with all of them. The runtime overhead of running multiple negotiations needs examination, as including more users might require additional user and ethics manager nodes.

\emph{Willingness to be Interrupted.} Initial considerations have been taken to model willingness to be interrupted based on practical factors, such as the estimated progress or effort of conflicting goals. This willingness could be included as an additional factor in ethical decision-making, where if the active user is less willing to be interrupted, the interrupting user enters negotiation with a small disadvantage, requiring higher ethical impact for precedence. These practical factors should not override the ethical concepts but only slightly influence decisions in specific situations. Measuring goal progress and modeling interruption willingness requires careful consideration. As a first step, the task achievements concept introduced in~\cite{Amigoni2015} could serve as a starting point by decomposing tasks into measurable milestones (e.g., reaching a location, grasping an object, etc.).

\emph{Strategic Behavior.} While the current design constrains opportunities for misrepresentation and deception through predefined and, where applicable, verified emergency categories, future work will investigate formal mechanisms for handling fully strategic or adversarial behavior. This includes incorporating uncertainty-aware reasoning over user-provided information, probabilistic modeling of unverifiable claims, and dynamic trust calibration mechanisms. Such extensions would enable the framework to operate robustly in settings where cooperative interaction cannot be assumed.
\vspace{-1.2em}
\section{Conclusion}
\label{sec:conclusion}
\noindent Public service robots increasingly operate in multi-user environments where human interruptions can create conflicts with ongoing tasks. Existing research has largely focused on efficiency and safety; in contrast, we introduced an \emph{ethical perspective}. 
Building on the idea of negotiation, we propose \emph{self-negotiation}, an intra-robot mechanism that allows a single robot to internally mediate between users’ ethical profiles. 
We presented a modular ROS-based architecture that realizes this framework and demonstrated its feasibility through simulations. The system consistently reached agreements aligned with user profiles and operated within acceptable bounds ($\sim$1--1.5s), demonstrating that ethical reasoning can be integrated into practical robot control.
Looking ahead, this work opens several avenues: deploying on real robots in public settings, scaling to multilateral conflicts and robot fleets, and incorporating practical factors such as task progress or effort. More broadly, the self-negotiation framework contributes to embedding ethical reasoning into everyday service robots, moving closer to autonomous systems that are not only safe and efficient but also aligned with human values.




\vspace{-0.25em}

\bibliographystyle{ieeetr}
\bibliography{ref}

\end{document}